\documentclass[sigconf,nonacm]{acmart}

\begin{document}
	
	\title{Laying foundations to quantify the "Effort of Reproducibility"}

	\author{Akhil Pandey Akella}
	\affiliation{%
		\institution{Northern Illinois University}
		\city{Dekalb}
		\state{Illinois}
		\country{USA}
	}
	\email{aakella@niu.edu}
	
	\author{David Koop}
	\affiliation{%
		\institution{Northern Illinois University}
		\city{Dekalb}
		\state{Illinois}
		\country{USA}
	}
	\email{dakoop@niu.edu}
	
	\author{Hamed Alhoori}
	\affiliation{%
		\institution{Northern Illinois University}
		\city{Dekalb}
		\state{Illinois}
		\country{USA}
	}
	\email{alhoori@niu.edu}

	\renewcommand{\shortauthors}{AP Akella et al.}
	
	\begin{abstract}
		Why are some research studies easy to reproduce while others are difficult? Casting doubt on the accuracy of scientific work is not fruitful, especially when an individual researcher cannot reproduce the claims made in the paper. There could be many subjective reasons behind the inability to reproduce a scientific paper. The field of Machine Learning (ML) faces a reproducibility crisis, and surveying a portion of published articles has resulted in a group realization that although sharing code repositories would be appreciable, code bases are not the end all be all for determining the reproducibility of an article. Various parties involved in the publication process have come forward to address the reproducibility crisis and solutions such as badging articles as reproducible, reproducibility checklists at conferences (\textit{NeurIPS, ICML, ICLR, etc.}), and sharing artifacts on \textit{OpenReview} come across as promising solutions to the core problem. The breadth of literature on reproducibility focuses on measures required to avoid ir-reproducibility, and there is not much research into the effort behind reproducing these articles. In this paper, we investigate the factors that contribute to the easiness and difficulty of reproducing previously published studies and report on the foundational framework to quantify effort of reproducibility.
		
	\end{abstract}

	
	\maketitle
	
	\section{Introduction}
	
	Scientific inquiry is established upon the pillars of community-driven reinforcing actions to establish confidence in published research \cite{NAP2019}. In order to foster a community that relies on refined levels of confidence and trust in scientific inquiry, a methodology must exist for validating this scientific rigor and certainty in results. Estimating the effort of reproducibility can be considered a fundamental unit of assessing scientific inquiry amongst published research. Current literature on terminologies associated with reproducibility \cite{Barba2018TerminologiesFR, Peng2011, stodden2014implementing} fall short of capturing the full spectrum of signals capable of encapsulating the effort behind reproducing a scientific work. \textit{Effort of reproducibility} is an important area worth exploring because it coalesces human cost, scientific validity, and confidence levels in scientific rigor into an impactful concept. Its existence, therefore, serves a greater good in scientific communities and is helpful in picking broad-spectrum signals when discussing cost, validity, and confidence associated with efforts to reproduce existing scientific work.
	
	When researchers across the social sciences sounded distress signals under the broader cause of \textit{reproducibility crisis \cite{baker2016reproducibility, hutson2018artificial, stupple2019reproducibility}}, it set forth momentum for researchers across various computational science disciplines to self-introspect and assess the extent of this reproducibility crisis. Additionally, entities such as journals, conferences, and academic and peer-review communities started taking note of the crisis, and released protocols \cite{boisvert2016incentivizing}, policies \cite{stodden2018empirical}, and checklists \cite{pineau2021improving}. To an extent, these measures of caution can safeguard future research from a range of reproducibility-related issues. Moreover, having checks and balances to foster a culture of reproducible research can sustain the trust and confidence the general public places in science. However, these actions cannot do justice to capture the human effort behind the journey of reproducibility. More importantly, if there exists a rubric to encapsulate the effort of reproducibility, we could benefit from including it in our existing discussions of checklists, protocols, and policies for ensuring reproducibility.
	
	Large-scale efforts such as the \textit{Reproduciblity Project \cite{open2012open} and Open Science Collaboration \cite{open2015estimating}} were established to overcome the disincentives of reproducing already published work. Furthermore, the existence of open-access peer-reviewed journals such as \textit{ReScience} \cite{rougier2017sustainable} provided a platform to critically analyze and voice concerns when replicating already published scientific papers along with the added incentive of publishing these replication reports. These large-scale projects and open-access journals are necessary for modern science as an institution to produce reproducible research. Given our current understanding of the refined scientific process, the existence of various reproducibility checklists, and the presence of a vibrant open-access peer-review community, we have several useful ingredients for discovering answers about the effort of reproducibility.
	
	Outlining these principles and unifying them under the framework of effort helps us contextualize the challenges associated with quantifying the effort itself. In this study, we lay a foundational groundwork for analyzing the effort of reproducing scientific articles from the research community's perspective. Data from the Machine Learning Reproducibility Challenge was collected and used for performing a.) an inductive qualitative analysis, b.) a quantitative analysis using Topic Models. The goal here is to systematically discover a distribution of factors responsible for encapsulating the effort of reproducibility.

	\section{Related Literature}
	The motivation to quantify the state of reproducibility in computational science research \cite{gundersen2018state} allowed critical dissection of the scientific method. The extent to which science is reproducible is a fundamental question in many studies such as \cite{raff2019step, raff2021research, yang2020estimating}. Although these studies might interchangeably use the word \textit{replicability} over \textit{reproducibility} \cite{Barba2018TerminologiesFR}, there is a higher level of agreement on the merit of quantifying reproducibility. But conceptually, reproducibility as highlighted by\cite{gundersen2018state}, is ``the ability of an independent research team to produce the same results using the same method based on the documentation made by the original research team."  
	
	Elaborate literature on reproducibility has provided members of the scientific community enough room to have a nuanced understanding of the terms and definitions associated with reproducibility. The new frontiers for reproducibility lie in doing something actionable with this knowledge. For this reason, we can observe plenty of interest \cite{salsabil2022study, wang2022res, rajtmajer2022synthetic} towards discovering, quantifying, and predicting research reproducibility at different levels such as \textit{Methods reproducibility and Results reproducibility} \cite{gundersen2018state}. Although these are significant endeavors, it is essential to discover the causal factors that contributed to the effort required to reproduce previously published work.
	
	Our work incorporates the motivations of all of the above mentioned studies to discover reasons responsible for easing or complicating the effort of reproducibility. Additionally, we showcase the benefit of taking inductive qualitative analysis (human) and combining it with quantitative topic models (machine) feedback. Essentially, knowledge from this approach can be incorporated as priors for various downstream modeling tasks from human-in-the-loop machine learning techniques \cite{wu2022survey}.
	
	\section{Building the Dataset}
	The Machine Learning Reproducibility Challenge from the years 2020 \cite{MLRC2020} and 2021 \cite{MLRC2021} provided a path for us to ask questions about the underlying effort in reproducing scientific articles. The primary goal of the ML Reproducibility Challenge is to have a community of researchers investigate the claims made in scholarly articles published at top conferences. The community selected papers and attempted to verify the claims made in the paper by reproducing computational experiments. The subsequent reports highlighting detailed information about the scope of reproducibility and what was easy and difficult for the researchers while replicating the original article were published on ReScience\footnote{http://rescience.github.io/read/} \cite{ReScience}. 
	
	ReScience is an open-access peer-reviewed journal with the goal of publishing researchers' endeavors to replicate computations of already published research using independently created, free, open-source software (FOSS). Demonstrably, the expected outcome from repeating experiments from previously published research is a verifiable status of reproducibility. Analyzing the additional information in these reports could support the research community in understanding the operational framework of effort necessary to reproduce published work.
	
	Although the first three versions \textit{ICLR`18, ICLR`19, NeurIPS`19} of the reproducibility challenge existed before 2020, we were interested in gathering data only from the recent editions of the ML Reproducibility Challenge (2020, 2021). The recent editions had reproducibility reports which followed a templated structure. The homogeneous nature of these reports was a motivating factor in focusing our analysis on recent editions. The combined articles from both editions (2020, 2021) were 87, of which 15 were removed because they belonged to adjacent disciplines of ML but not ML itself. Additionally, two more articles were removed from the final dataset because they were editorials of the reproducibility challenge. The final tally of articles selected from the reproducibility challenge is 70. We also got information about the original articles (i.e., authors and meta-level) from Google Scholar. 
	We built scripts to automatically extract this information using python software libraries such as urllib3, Beautiful Soup, and BibtexParser. The preliminary dataset we constructed after systematic data collection consisted of the following aspects:
	
	\begin{enumerate}
		\item \textbf{Meta information about the reproducibility report} such as author's name, title, DOI, year, volume, and issue.
		\item \textbf{Meta information about the original work} such as author's name, title, DOI, year, and Google Scholar citations.
		\item \textbf{Digital artifacts} such as OpenReview submission of the reproducibility study, PDF of the reproducibility report, and code repository of the replication. 
	\end{enumerate}
	
	Our code repository with every artifact, data, and experiments is available on Github \footnote{https://github.com/reproducibilityproject/effortly}.
	
	\section{Inductive Qualitative Analysis}
	
	We choose an inductive approach for potentially establishing a relationship between the text seen in reproducibility reports to the concept of effort. For our use case, a general inductive approach would mean repetitive, structured reading of the reproducibility reports to discover categories of reasons that made it easier or difficult to reproduce original works and  reasons that served as limitations while evaluating the reproducibility of original works. The last component before initializing the analysis was having composite information of all the reproducibility reports (downloaded as PDFs) into a structured form. For this, we have utilized Allen-AI's Science Parse\footnote{https://github.com/allenai/science-parse} software package to parse all of the sections of the reproducibility report.
	
	\begin{figure}[h]
		\centering
		\includegraphics[width=\linewidth]{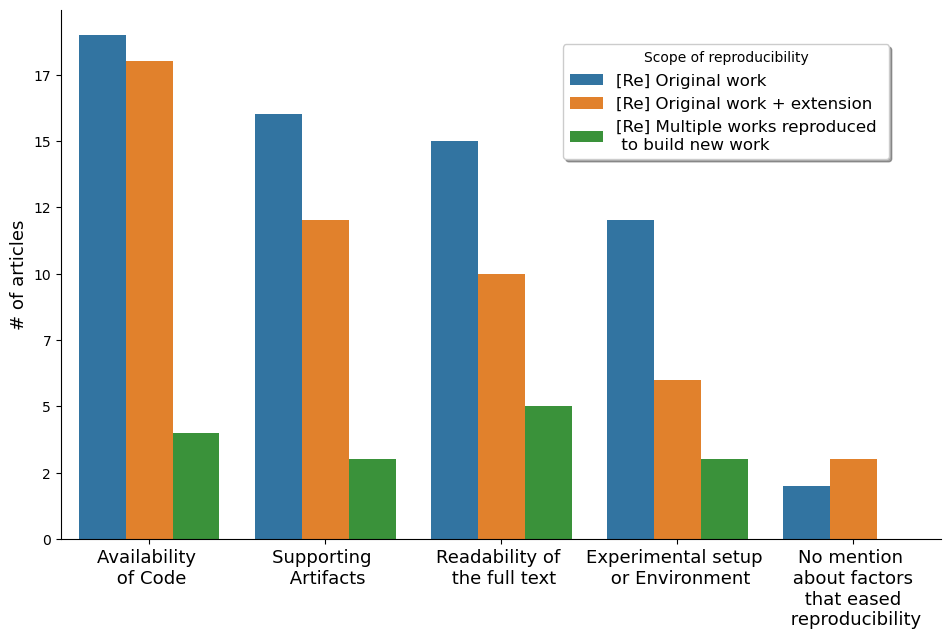}
		\caption{Reasons that eased the effort to reproduce.}
		\label{fig:reason-easy}
		\Description{"What was easy" while reproducing published work}
	\end{figure}
	
	\subsection{Latent variables to articulate effort}
	
	Observing a general pattern amongst the reproducibility reports would require encoding sections of the report as labeled information. Almost every report from the reproducibility challenge followed a template that included sections such as the scope of reproducibility and reasons for easiness and difficulty of reproducibility. The purpose behind including these sections was to allow the community to reflect on the individual researcher’s insight while reproducing the original article. These insights included critique on elements such as clarity, thoroughness, and correctness of the original article. Therefore, it was pertinent to encode this information from the aforementioned sections as they could be latent variables acting as pointers for generalizing the effort of reproducible articles. To that extent, we extracted three sections, “Scope of Reproducibility,” “What was easy,” and “What was difficult” to build these latent variables.
	
	\subsection{Scope of Reproducibility}
	“Scope of Reproducibility” is the first section we extracted from the reproducibility report. The section's purpose is to outline main contributions of the original paper , the specific setting or problem addressed in the paper, and list the experimental methodology adopted to solve the problem. Each claim should be relatively concise; some papers may not clearly list their claims, and one must formulate them in terms of the presented experiments. The claims are roughly the scientific hypotheses evaluated in the original work. Analytically, we observed three potential categories describing the attempts to reproduce original works.
	
	\begin{enumerate}
		\item \textbf{Original work}: Replications observed under this category are straightforward implementations of the original work.
		\item \textbf{Original work with supplemental extensions}: Replications of this category include implementations of the original work and additional contributions. An example of this category is the replication of the original work and the inclusion of an ablation study.
		\item \textbf{Reproducing more than one work to build new work}: Replications of this category include implementations of more than one original work to support a completely new idea.
	\end{enumerate}
	
	\subsection{Factors that made it easy or difficult to reproduce original study}
	The sections “What was easy” and “What was difficult” represent two sides of the same variable but are helpful in establishing the individual factor that is responsible for easing or burdening the effort to reproduce the research. Additionally, having two separately encoded variables (easy and difficult) highlights the existence of co-dependent factors. For instance, the availability of software artifacts might have made it easy for an individual researcher to initiate the process of reproducing a published article, but the lack of hyperparameters or data might put a strain on the effort behind reproducing the original article. Although checking for the availability of software artifacts and information about hyperparameters can be unified using the question "Is code available?", we can clearly see how the current situation cannot be boiled down into a "yes" or "no" answer.
	
\begin{figure}[h]
		\centering
		\includegraphics[width=\linewidth]{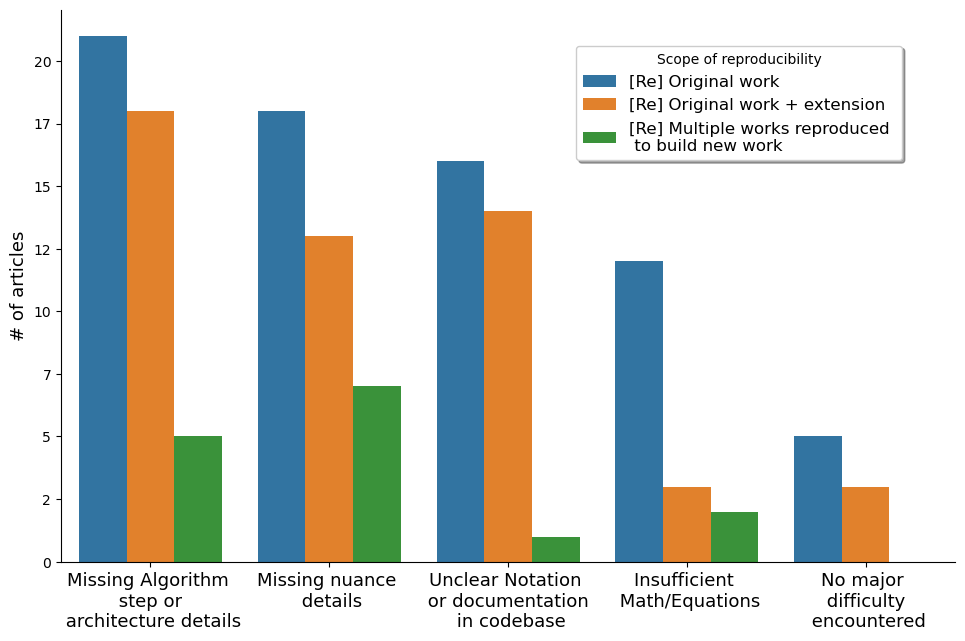}
		\caption{Reasons that made it difficult to reproduce.}
		\label{fig:reason-difficult}
		\Description{"What was difficult" while reproducing published work}
	\end{figure}

	A rigorous process of inductive encoding meant systematic reading and re-reading of the reproducibility reports to create descriptive categories that link associations to the reasons why a researcher found an original work easier or difficult to reproduce. Therefore, assumed causal reasons observed both in Fig \ref{fig:reason-easy} and Fig. \ref{fig:reason-difficult} are borne out of careful human consideration.
	
	A visual translation of our analysis can be observed in Fig \ref{fig:reason-easy}. We can observe that the distribution of factors that made it easier to reproduce original works include \textit{Availability of Code, Supporting Artifacts, Readability of Full text, Experimental setup or Environment}, and in some cases, \textit{No mention about factors that eased reproducibility}.
	
	Similarly, visual translation of causal reasons that made it difficult to reproduce original works can be observed in Fig \ref{fig:reason-difficult}. These factors include \textit{Missing Algorithm step or Architecture details, Missing nuance details, Unclear notation or documentation in the codebase, Insufficient Math/Equations}, and in some cases, \textit{No major difficulty encountered}.
	
	Identifying reasons for easiness or difficulty is a pursuit that is both contextual and subjective. Under \textbf{contextual reasons for easiness}, Fig. \ref{fig:reason-easy} suggests that the \textit{Availability of Code} and \textit{Supporting Artifacts} are the most important contextual factors that make it easier for a researcher to replicate original works. Similarly, \textbf{contextual reasons for difficulty}, Fig. \ref{fig:reason-difficult} show \textit{Missing Algorithm steps or Architectural details} along with \textit{Missing Nuance details} to be the most important factors that made it difficult for a researcher to replicate the original work.
	
	Under \textbf{subjective reasons for easiness or difficulty}, we can notice from Fig. \ref{fig:reason-easy} that the \textit{Readability of the full text} is a crucial subjective factor that eased the effort to reproduce the original study whilst under difficulty \textit{Unclear notation or documentation in the codebase} to be a vital subjective factor that made it difficult to replicate original work. Also, it is interesting to notice a \textbf{shift in priority over reasons for easiness or difficulty} for studies whose scope is \textit{Reproducing more than one work to build new work}. For instance, in Fig \ref{fig:reason-easy}, the \textit{Readability of the full text} is mentioned more as a reason than the most prevailing factor, the \textit{Availability of Code}. Adjacently, under reasons for difficulty, as seen in Fig \ref{fig:reason-difficult}, \textit{Missing nuance details} is the dominant factor. 
	
	\subsection{Factors that limited the evaluation of reproducibility}
	The purpose of discovering limitations within a reproducibility study is to indicate any potential factor(s) that served as a restriction to the researcher while  re-implementing the original methodology. The marginal impediments in the researcher's reproducibility journey are elemental to understanding the necessary measures taken by the researcher to overcome any limitations.
	
	Capturing this information from ReScience reports was a starting point to understand the underlying limitations while reproducing the machine learning papers because, many a time, reproducing a machine learning study means pooling large compute resources, choosing the right hyperparameters, etc. Although these limitations are minor hurdles while replicating the original work, they do not necessarily translate into reasons for irreproducibility. Although all 70 samples in our data are successful reproductions of the original study, there still exist reasons for limitations while evaluating the original study.
	
\begin{figure}[h]
		\centering
		\includegraphics[width=\linewidth]{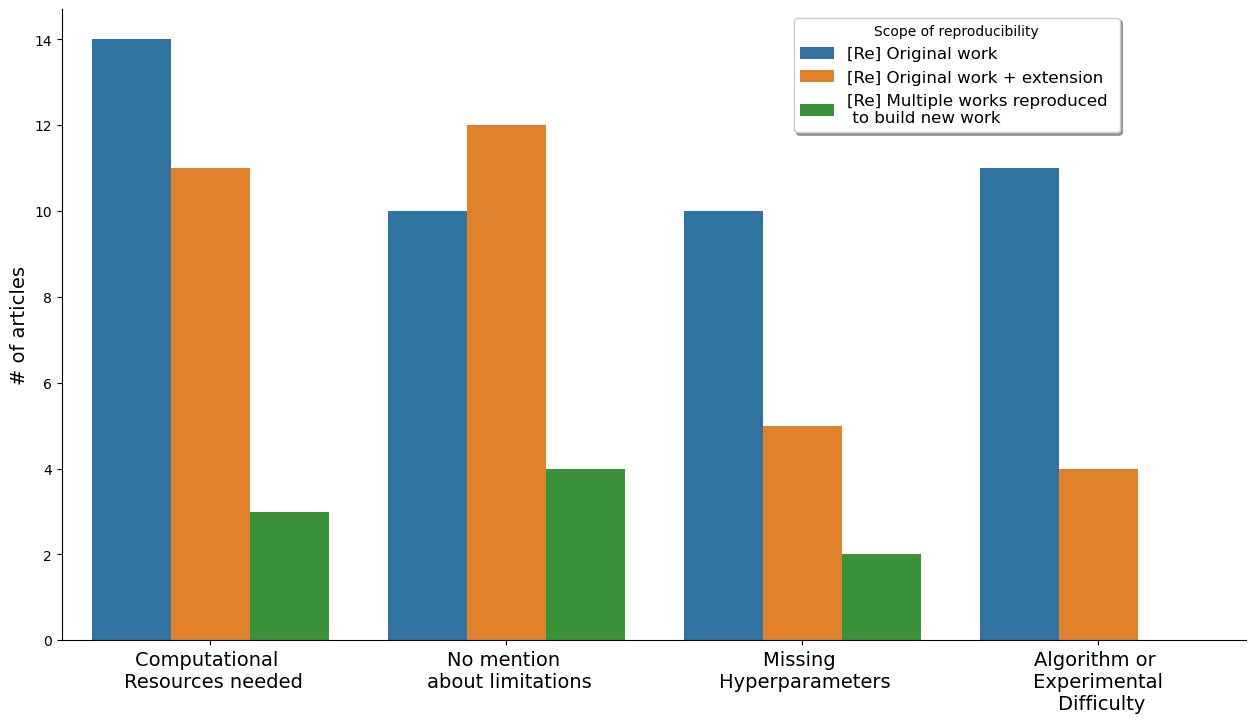}
		\caption{Reasons that served as limitations while evaluating the effort to reproduce.}
		\label{fig:reason-limitations}
		\Description{"Limitations" while reproducing published work}
	\end{figure}

	Fig. \ref{fig:reason-limitations} visualizes the distribution of factors that limited the evaluation of reproducibility, and it includes the \textit{Necessity of Computational Resources,  Missing Hyperparameters, Algorithm or Experimental Difficulty, or No mention about limitation}.
	
	The above mentioned shift in priorities for studies can be noticed in Fig \ref{fig:reason-limitations} as well. For instance, studies under the scope \textbf{Reproducing more than one work to build new work} never mention \textbf{Algorithm or Experimental difficulty} as a subjective factor for limitation whilst studies under the scope \textbf{Original work}, consider the exact same factor to be second most important reason for limiting the evaluation of reproducibility.
	
	\section{Topic Modeling}
	
	Topic Models help us discover latent variables semantically related to concepts that signal the effort required to reproduce articles. Utilizing the raw texts from each report helps us preserve the distinct properties of each document whilst providing a unique representation of the mixture of reasons related to effort. Latent Dirichlet Allocation (LDA) \cite{blei2003latent} is a popular way to perform topic modeling. We utilized LDA to model reproducibility reports since those documents consist of topics. Conceptually, this is relevant to our goal, as factors that make it easier or difficult to reproduce an original work could be present as random mixtures over latent topics. We built a corpus of documents $D_{easy}$, and $D_{difficult}$ consisting of raw texts of "What was easy" and "What was difficult" and trained two models $LDA_{easy}$, and 
	$LDA_{difficults}$. The topic coherence score for $LDA_{easy}$ is 0.415, and $LDA_{difficult}$ 0.326. We have used the Python library \textbf{GenSim} to build the LDA models. The optimum number of topics for the LDA model was decided by ranking topic coherence scores against the number of topics. 
	
	\begin{table}
		\caption{Most relevant terms observed by the LDA model when trained on "What was easy" corpus.}
		\label{tab:easy-lda}
		\begin{tabular}{cc}
			\toprule
			Topic & Most relevant terms\\
			\midrule
			1 & describe, straightforward, understand, documented\\
			2 & codebase, repository, source, instructions\\
			3 & datasets, training, scripts, experiments\\
			4 & results, ideas, evaluation, architecture\\
			5 & correspondence, addressed, peer-review, copyright\\
			\bottomrule
		\end{tabular}
	\end{table}
	
	Table \ref{tab:easy-lda}, and Table \ref{tab:difficult-lda} showcase the most relevant terms observed by the LDA model when trained on their respective corpora. These terms are obtained by querying the respective topics for most dominant keywords by percentile contribution. We set a threshold of greater than 0.9 to notice the most relevant topic keywords. Analyzing the relevant terms in Table \ref{tab:easy-lda}, we can notice close similarity towards factors mentioned in Fig \ref{fig:reason-easy}. This outcome reinforces the relevance of our inductive encoding strategy. Interestingly, the importance of including a topic model in our study can be noticed after observing Topic five from Table \ref{fig:reason-easy}. Thematically, it suggests \textbf{Author Correspondence} to be a dominant topic cluster observed by the LDA model as a relevant factor from "What was easy" corpus. Therefore, communication with the authors can be considered as an important factor that can ease the effort of reprocibility. 
	
	\begin{table}
		\caption{Most relevant terms observed by the LDA model when trained on "What was difficult" corpus.}
		\label{tab:difficult-lda}
		\begin{tabular}{cc}
			\toprule
			Topic & Most relevant terms\\
			\midrule
			1 & dataset, algorithm, implementation, method\\
			2 & training, loss, accuracy, learning\\
			3 & models, network, training, time\\
			4 & difficult, challenges, evaluation, claim\\
			5 & methods, features, performance, parameters\\
			\bottomrule
		\end{tabular}
	\end{table}

	\section{Conclusion}
	In this study, we lay the foundations to analyze and discover factors that can encapsulate the effort of reproducing scientific articles. Our approach of combining inductive qualitative analysis with topic modeling resulted in discovering factors that serve as reasons for easiness, reasons for difficulty, and limitations while evaluating the reproducibility of previously published work.
	
	\section{Acknowledgement}
	This work is supported in part by NSF Grant No. 2022443.
	
	
	\bibliographystyle{ACM-Reference-Format}
	\bibliography{effortly}

\end{document}